
\input harvmac

\Title{\vbox{\baselineskip12pt \hbox{HUTP-92/A038}
\hbox{CERN-TH 6608/92}\hbox{August 1992}  } }
{Neutrino Helioseismology}
\centerline{A. De R\'ujula$^{*\star}$ and S.L. Glashow$^{\ddagger\star}$}

\vskip .25in
\centerline{ $*)$ CERN, 1211 Geneva 23 (permanent address)}

\centerline{ $\ddagger)$ Physics Department, Harvard
University, Cambridge, MA 02138}

\centerline{ $\star)$ Physics Department,
Boston University, Boston, MA 02215}
\vskip .4in
\parskip=2pt

\centerline{Abstract}
\noindent The observed deficit of $\rm ^8B$ solar neutrinos may
call for an improved standard model of the sun or an expanded
standard model of particle physics ({\it e.g.,} with
neutrino masses and mixing). In the former
case, contemporary  fluid motions and thermal fluctuations in the
sun's core may
modify nuclear reaction rates and restore agreement.  To test this notion,
we propose a search for short--term variations of the solar neutrino flux.

\Date{CERN-TH 6608/92, HUTP-92/A038 }

{\bf
The observed deficit of $\rm ^8B$ solar neutrinos
}
\ref\rdavis{Davis Jr., R. {\it et al.} in {\it Proceedings of the 21st
International Cosmic Ray Conference} (ed Protheroe, R.J.) 155--158
(University
of Adelaide Press, Adelaide 1990).}%
\nref\rkam{Hirata, K.S. {\it et al.,} Phys. Rev. {\bf D44,}
2241--2260 (1991).}%
\nref\rga{GALLEX Collaboration (Anselman, P. {\it et al.,}) Phys. Lett.
{\bf B285,} 376--389 (1992).}%
--\ref\rsage{Abazov, A.I. {\it et al.,} Phys. Rev. Lett. {\bf 67,}
3332--3335 (1991).}
{\bf
 may call for an improved standard model of the sun
 or an expanded
 standard model of particle physics ({\it e.g.,} with
neutrino masses and mixing). In the former
case, contemporary  fluid motions and thermal fluctuations in the
sun's core may
modify nuclear reaction rates and restore agreement
}
\ref\rrox{Roxburgh, I.W. in {\it The Internal Solar Angular Velocity}
(ed Durney, B.R. \& Sofia, S.) 1--5 (Reidel, Dordrecht 1987).}
\ref\rgougha{Gough, D.O.,  Ann. N.Y. Acad. Sci. (in press).}.
{\bf
To test this notion, we propose a search for
short--term variations of the solar neutrino flux.
}

Models of the sun fit its radius $R_{_\odot}$ and luminosity $L_{_\odot}$
to an assumed
initial ${\rm ^4He}$ abundance and a convective
mixing length
\ref\rschw{Schwarzchild, M., Howard, R. \& H\"arm, R.,
Astroph. J. {\bf 125,} 233--241 (1957).}%
\nref\rbah{Bahcall, J.N.
\& Ulrich, R.K.,  Rev. Mod. Phys. {\bf 60,} 297-372  (1988).}%
\nref\rturk{Turk--Chi\`eze, S. in {\it Inside the Sun} (eds
Berthomieu, G. \&
Cribier, M.) 125--132 (Kluwer Acad. Pub., Dordrecht 1990).}%
--\ref\rchris{Christensen--Dalsgaard, J., Gough, D.O. \&  Thompson,
 M.J., Astroph. J. {\bf 378,} 413--437 (1991).}.
While challenged by solar--neutrino observations,
they are supported by solar-surface measurements
\ref\ehse{Duvall Jr., T. L. in {\it Inside the Sun,}
{\it op. cit.}  253-264.}
 of the frequencies of thousands of $p$--waves (pressure waves).
These are inverted to yield the sound velocity at depth
\ref\rinva{Gough, D.O.,  Solar Physics {\bf 100,} 65--99 (1985).}
\ref\rinvb{Gough, D.O. \& Thompson, M.J.
in {\it Solar Interior and Atmosphere,} (ed
Cox, A.N., Livingstone, W.C.  \&
Mathews, M.) 401--478
(Space Science Series, University
of Arizona Press, Tuscon 1991).}.
Whilst the result agrees with solar models to
better than 1\%,
helioseismology provides scant information about the
solar core, where $p$ waves are damped
\ref\rhst{Christensen--Dalsgaard, J. \& Berthomieu, G.
in {\it Solar Interior and Atmosphere, ibid.} 519-561.}.

Solar $g$ waves (for which gravity is the restoring force)
are suppressed toward the surface and difficult to see,
but they may well be present. As the sun evolves, the
$\rm ^3He$ abundance in its core develops a
positive outward gradient. This leads
\ref\rgwave{Dilke, F.W.W. \& Gough, D.O.,
Nature {\bf 240,} 262--264 cont'd
293--294 (1972).}
\ref\rhwave{Merryfield, W.J., Toomre J. \&  Gough, D.O.,
Astroph. J. {\bf 367,} 658--665 (1991).}
to a hydrostatic instability
(often ignored in standard solar models)  and to the  secular growth of
radially asymmetric standing $g$ waves of low order $n$
(number of radial nodes) and degree $l$ (multipole moment).
Their periods $(2\pi/\omega)$ are of order one hour \rhst\
\ref\rhill{Hill, H. {\it et al.}
in {\it Solar Interior and Atmosphere, op. cit.} 562--617.}.
Since energy--tranport times are much larger,
$g$ waves correspond to quasi--adiabatic temperature fluctuations
about a radial mean:
\eqn\gamp{T(r,\,t,\,\theta,\,\phi)={\overline T}(r)\left[
1+A\,g(r)\,Y_{lm}(\theta,\,\phi)
\,\sqrt{2}\,\cos(\omega t)\right],  }
where $A$ is the amplitude of an  oscillation whose
angular dependence is that of a spherical
harmonic  with $\int \vert Y\vert^2 d\Omega =1$ and whose
radial eigenfunction $g(r)$ has a maximum of one.

Any $g$ wave present in the sun affects its neutrino--producing
processes:
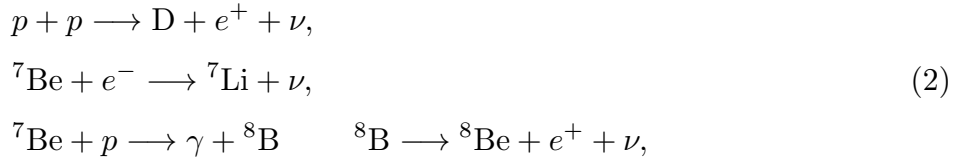
\eqn\ereact{\eqalign{
&p+p \longrightarrow {\rm D} +e^++\nu,   \cr
&{\rm ^7 Be} + e^- \longrightarrow {\rm ^7Li} + \nu, \cr
&{\rm ^7 Be} + p \longrightarrow \gamma+ {\rm ^8B} \qquad\
 {\rm ^8B}
\longrightarrow {\rm ^8Be}+ e^+ + \nu, \cr  }}
which we label $a=1,\,7,\,8$.
Their angularly--averaged rates $\hat{\epsilon}_a(r,t)$
are:
\eqn\epsi{
\hat{\epsilon}_a(r,t)=
\epsilon_{a}(r)\,
\langle\, (T/\overline T)^{N_a}\,  \rangle_\Omega,}
where the
$\epsilon_{a}(r)$  depend on the local density, nuclear abundances
and $\overline{T}(r)$.
The exponents in \epsi,
$N_1=4,\ N_7\simeq -0.5\ {\rm and\ } N_8\simeq 13,$
reflect the $T$ dependences of the
reaction rates at fixed abundances \rgougha.
Expanding \epsi\ in powers of $A^2$,
we exhibit the time dependence of the rates:
\eqn\epsil{ \hat{\epsilon}_a(r,t)=
\epsilon_{a}(r)\,\left( 1+ {1\over2}\,A^2\,N_a(N_a-1)\,g^2(r)\,
[1+\cos(2\omega t)] + O(A^4)\right).}
The constant in square brackets affects the time--averaged neutrino
fluxes; the cosine generates oscillations with
twice the $g$ wave frequency.

We integrate \epsil\ over $r$ using an $n=1$ mode with  $g(r)=x \exp(1-x)$,
$x=r/(0.15\,R_{_\odot})$ \rgougha. In solar models \rbah--\rchris,
$\epsilon_{a}(r)$ are roughly of the form $\epsilon_{a}(r)=y^2\,\exp(-y^2)$
with $y=r/(f_a\,R_{_\odot})$ and $f_a\simeq 25,\, 17,\, 10$ for
$a=1,\, 7,\, 8$.
For the oscillations of the various components of the solar neutrino
flux, we predict:
\eqn\nuosc{
F_a \simeq \overline{F}_a\,[ 1+ \lambda_a\,\cos(2\omega t)],
\qquad\ \lambda_{1,7,8}\simeq(4.8,\,0.21,\,28.5)\,A^2,}
where $\overline{F}_a$ are
time--averaged fluxes. Notice that $\lambda_8> \lambda_1 >\!> \lambda_7$.

According to \epsi,
the reaction rates ${\hat \epsilon}_a$ exceed those
in a steady sun with temperature profile $\overline{T}(r)$. To keep
$L_{_\odot}$ fixed, the solar model must be modified to lower
$\overline{T}$. Gough \rgougha\ estimates how the
time--averaged neutrino fluxes depart, in the presence of an $n=1$
$g$-wave, from those of the standard
model:\footnote{$^1$}{The burning of H to ${\rm ^4He}$
is the main source of solar energy, so that
${F}_1$ is hardly affected.
${F}_7$ and ${F}_8$ are suppresed \rrox,
 since the effects of  reducing
$\overline{T}$ win over the enhancement obtained from the
time average of \epsil.}
\eqn\caca{ \eqalign{
\overline{F}_1 &\simeq  F_1^{SSM}, \cr
\overline{F}_7 &\simeq  F_7^{SSM}\,  [(1-33\,A^2 + 267\,A^4], \cr
\overline{F}_8 &\simeq  F_8^{SSM}\,  [1-57\,A^2 + 1067\,A^4].\cr } }
The effects of the $g$ wave on the time--averaged flux \caca\ and its
fluctuations \nuosc\ are greatest for $\rm^8B$ neutrinos. With
$A=0.1$, the $\rm ^8B$ flux is reduced by 0.54 and that of
$\rm ^7Be$ by 0.70, removing
the discrepancy between experiment and theory. We choose this value of $A$
to set the scale for anticipated neutrino flux oscillations.

Future experiments will measure
arrival times $t_i$ of thousands of neutrinos.
Assume that a $g$ mode of frequency $\omega$ modulates
the neutrino fluxes, as in \nuosc.
The precise frequency of the
$g$ wave is unknown, but
its effect can be found by Fourier
transforming the data over a frequency range
$f_{{\rm min}}<f<
 f_{\rm max}$.
Suppose that $n$ neutrinos are
detected in a run of duration $\tau$.  Let:
\eqn\ep{P(f)\equiv\left\vert \sum_{j=1}^{n} e^{ift_j} \right\vert  .  }
The signature of a $g$ wave is
a peak in $P(f)$ at $f=2\omega$, emerging even if $2\omega$
exceeds $\tau/n$,  the mean counting
rate. The peak's
expected magnitude is
$P_s=\lambda\,n/2$.  Its half--width at half maximum,
$\Delta \omega= \sqrt{6}/\tau$, sets the required Fourier resolution.
Away from the peak, $P(f)$ fluctuates about
$\sqrt{n}$, exceeding $P_s$ with probability
$\exp(-P_s^2/2n)$.

The minimum significant signal (with confidence level C.L.)
corresponds to a $g$ wave of amplitude:
\eqn\eftr{\lambda_{{\rm min}}=\left[{8\over n}\;
\ln{(f_{\rm max}-f_{{\rm min}})\,\tau\over
1-{\rm C.L.}}\right]^{1/2},   }
With $\tau \sim 1\;\rm  year $ and $f_{\rm max}-f_{\rm min}\sim$
inverse minutes, the  logarithm's argument
is large and its precise value
immaterial.

We see from \nuosc\ and \eftr\ that
for $A\simeq 0.1$,  an experiment sensitive to the $pp$ flux
must observe
$\sim\!6\times10^4$  events to find a 99\%-confidence effect.
A real--time $pp$
neutrino detector with this capability has been discussed (Ypsilantis,
T.  \& Seguinot, J., priv. com.).
The proposed BOREX experiment (Raghavan, R. {\it
et al.,} Bell Laboratory Report No. ATT-BX-88-01 (1988)) could detect
a million ${\rm ^7Be}$ neutrinos, but falls
short of the $\sim\!3\times10^7$ events
needed to detect the tiny oscillations expected in this case.

Fewer events suffice to detect oscillations of the $\rm ^8B$
neutrino flux.
The Sudbury Neutrino Observatory
\ref\rsno{Beier, E. in {\it Proc. of the International Symposium on
Underground Physics Experiments} (ed Nakamura, K.) 165--170
(Institute for Cosmic Ray Research,
University of Tokyo 1990).},
Super--Kamiokande
\ref\rsk{Totsuka, Y., {\it ibid.}  129--164.}
and Icarus
\ref\ric{Baldo--Ceolin, M.  in {\it Massive Neutrinos in Particle Physics and
Astrophysics} (ed Fackler, O.W. \& Tran Thanh Van, J.) 159--164
(Editions Fronti\`eres, Gif sur Yvette 1986).}
each will time thousands of these neutrinos. We deduce from \nuosc\ and \eftr\
that an  experiment gathering
 3000 (30,000) events can find an $A=0.08\;(0.05)$
signal with 99\% confidence, decisively testing  whether the suppression of
the ${\rm ^8B}$ neutrino flux is due to a single
$g$--mode.%
\footnote{$^2$}{If the $\rm ^8B$ neutrino deficit results
from several $g$ modes rather
than a dominant one,  their Fourier powers are smaller and  a
model--independent
search becomes more difficult.}

If neutrino experiments were to detect the sun's heartbeat,
it is the sun that oscillates, not the neutrino.

\bigbreak\bigskip\bigskip\centerline{{\bf Acknowledgements}}\nobreak
We thank John N. Bahcall and Douglas O. Gough
for helpful discussions.
This work was supported in part by NSF contract PHY87--14654.

\listrefs

\bye

FOOTNOTE ABOUT MSW OSC DUE TO EARTH
(we do not dwell on the possibility of relating weak
candidate peaks with the help of \eone.

 The value $A\sim 0.1$ indicated by the overall ${\rm ^8B}$ flux
suppresion can be established in an ${\rm ^8B}$ experiment with a mere 2000
events. For a $pp$ experiment to reach the same sensitivity, a total of
events would be required, which is not out of the question.

Suppose that several $g$ modes are excited, so that the neutrino flux is

modulated by $1+\sum \lambda_i\,\cos(\omega t)$. If no one $\lambda$
exceeds the threshold for detection, the Fourier transform
cannot establish their presence and another technology must be brought to
bear.

If their frequencies are $\omega_i$, the $\rm ^8B$ neutrino flux, and
hence the probability of their detection, is modulated by a factor of the
form $1+\sum \lambda_i \cos(2\omega_i t)$.

Gough and Gratz point out [7] that the standard solar models are unstable.
Fluid motions in the solar core  can lead to a reduction of the
higher-energy neutrino flux [8].  We propose a search for short--term
fluctuations of the intensity of the solar neutrino flux, whose oscillatory
or chaotic behavior would show that the sun's core is not in a steady
state.

strengthened by ing that
steady--state calculations overestimate the flux of the high energy
$\rm ^8B$
component.  Time dependent variations of the temperature  of the sun's core

 We adopt this point of view and propose a
search for detectable short--term variations of the solar neutrino flux.

\bye